\documentclass{aa}  
\usepackage{graphicx}
\usepackage{txfonts}
\usepackage{natbib}
\bibpunct{(}{)}{;}{a}{}{,} 
\begin{document}

\title{SPH simulations of grain growth in protoplanetary disks}

\author{G. Laibe\inst{1} \and J.-F. Gonzalez\inst{1} \and L. Fouchet \inst{2}
\and S.T. Maddison\inst{3}}


\institute{Universit\'e de Lyon, Lyon, F-69003, France;
Universit\'e Lyon 1, Villeurbanne, F-69622, France;
CNRS, UMR 5574, Centre de Recherche Astrophysique de Lyon;
\'Ecole Normale Sup\'erieure de Lyon, 46 all\'ee d'Italie,
F-69364 Lyon cedex 07, France\\
\email{[Guillaume.Laibe;Jean-Francois.Gonzalez]@ens-lyon.fr}
\and
Institute of Astronomy, ETH Z\"urich, Schafmattstrasse 16, HPT D19,
CH-8093 Z\"urich, Switzerland\\
\email{fouchet@phys.ethz.ch}
\and
Centre for Astrophysics and Supercomputing, Swinburne Institute of Technology,
PO Box 218, Hawthorn, VIC 3122, Australia
\email{smaddison@swin.edu.au}}

\date{Received 6 February 2008 / Accepted 5 June 2008}

\abstract
{}
{In order to understand the first stages of planet formation, when tiny grains
aggregate to form planetesimals, one needs to simultaneously model grain
growth, vertical settling and radial migration of dust in protoplanetary disks.
In this study, we implement an analytical prescription for grain growth into a
3D two-phase hydrodynamics code to understand its effects on the dust
distribution in disks.}
{Following the analytic derivation of Stepinski \& Valageas (1997), which
assumes that grains stick perfectly upon collision, we implement a convenient
and fast method of following grain growth in our 3D, two-phase (gas+dust) SPH
code. We then follow the evolution of the size and spatial distribution of a
dust population in a classical T~Tauri star disk.}
{We find that the grains go through various stages of growth due to the
complex interplay between gas drag, dust dynamics, and growth. Grains
initially grow rapidly as they settle to the mid-plane, then experience a fast
radial migration with little growth through the bulk of the disk, and finally
pile-up in the inner disk where they grow more efficiently. This results in a
bimodal distribution of grain sizes. Using this simple prescription of grain
growth, we find that grains reach decimetric sizes in $10^5$~years in the
inner disk and survive the fast migration phase.}
{}


\maketitle

\section{Introduction}
\label{Intro}

The first steps of planet formation are governed by the build-up of
planetesimals due to the dust coagulation in protoplanetary disks
\citep{Dominik2007}. Observational evidence for grain growth in disks is now
common \citep{Apai2004,Rodmann2006,Muzerolle2006,Lommen2007,Graham2007}. 
Grains must grow from sub-$\mu$m sizes to planetesimal scale (kilometer size)
objects in a fraction of the lifetime of the disk, which is estimated to be a
few $10^7$ years \citep{Haisch2001,Carpenter2005}. The timescales of grain
growth, however, are unclear: some young disks show signatures of grain growth
while old disks can show signatures of unprocessed grains and coeval disks can
show a range of grain sizes and dust processing \citep{Kessler-Silacci2006}.

Grains can grow via collisions and depending on their relative velocity
and on their chemical and physical properties \citep{Chokshi1993,Blum2006},
colliding grains can rebound, shatter or stick. Grains will settle vertically
and migrate radially at different rates according to their size
\citep{Weidenschilling1977,Garaud2004,BF05}, leading to local density
enhancements in the disk. Since grain growth is dependent on density, changes
in the dust distribution will affect growth rates, which in turn will affect
the dynamics of the dust \citep{Weidenschilling1980,Haghi2005}. Therefore,
growth, settling and migration need to be simulated together.

Various models have been developed to describe the grain growth process. One
approach is to use the time-dependent Smoluchowski coagulation equation
\citep{Weidenschilling1980,Weidenschilling1997,SuttnerYorke2001,
DullemondDominik2005,Tanaka2005,Nomura2006,Ciesla2007} which describes the
number density evolution of particles of a given mass range. The numerical
solution of the Smoluchowski equation is challenging. Another approach is to
use an analytic expression for the grain growth rate as a function of local
disk conditions \citep{Stepinski1997,Haghi2005}. In this study, we use this
second approach and implement the analytical prescription of
\citet[hereafter SV97]{Stepinski1997}\defcitealias{Stepinski1997}{SV97}
in our 3D, two-phase (gas+dust) hydrodynamics code and follow the evolution of
the grain size distribution in a protoplanetary disk. We validate our method
on an axisymmetric disk here, before applying it to non-axisymmetric complex
problems in future work.

\section{An analytical grain growth model}
\label{SectSV97}

\citetalias{Stepinski1997} modeled the radial evolution of solid particles
made of water ice in geometrically thin, turbulent, vertically isothermal
protoplanetary disks. They described the gas and solid particle components as
two separate phases coupled by aerodynamic forces (in the Epstein regime), and
assumed that the evolution of the gas is unaffected by that of the solids. The 
particles can grow by coagulation (they stick perfectly upon collision and 
therefore never shatter into smaller grains), evaporate or condense from vapor.
Their size distribution at any radius and time is supposed to be narrowly
peaked around a local mean value $s(r,t)$. Self-gravity is neglected.

They obtained an analytic expression for the evolution of the particle size
$s$ given by
\begin{equation}
\frac{\mathrm{d}s}{\mathrm{d}t}=\sqrt{2^{^3\!/\!_2}\,\mathrm{Ro}\,\alpha}\,
\frac{\hat{\rho}_\mathrm{d}}{\rho_\mathrm{d}}\,C_\mathrm{s}
\frac{\sqrt{\mathrm{Sc}-1}}{\mathrm{Sc}},
\label{EqEvol}
\end{equation}
where Ro is the Rossby number for turbulent motions, $\alpha$ the
\citet{ShakuraSunyaev1973} viscosity parameter, $\hat{\rho}_\mathrm{d}$ the
density of matter concentrated into solid particles, $\rho_\mathrm{d}$ the
bulk density of the grains, $C_\mathrm{s}$ the local gas sound speed,
and Sc the Schmidt number of the flow which estimates the effect of gas
turbulence on the grains. They defined Sc as
\begin{equation}
\mathrm{Sc}=(1+\Omega_\mathrm{K}\,t_\mathrm{s})
\sqrt{1+\frac{\bar{\vec{v}}^2}{V_\mathrm{t}^{2}}},
\label{EqSchmidt}
\end{equation}
where $\Omega_\mathrm{K}$ is the local keplerian velocity, $t_\mathrm{s}$
the dust stopping time, $\bar{\vec{v}}$ the mean relative velocity between
gas and dust, and $V_\mathrm{t}$ a turbulent velocity. Note that
\citet{Youdin2007} recently suggested that one should use
Sc~$\simeq 1+(\Omega_Kt_s)^2$ instead. However it is not clear how this
approximate value follows from their more complex exact expression
$(1+(\Omega_Kt_s)^2)^2/(1+4(\Omega_Kt_s)^2)$. In this paper, we keep the
formulation used by SV97 for the sake of self-consistency.

The growth rate $\displaystyle\frac{\mathrm{d}s}{\mathrm{d}t}$ depends on $s$
via the stopping time
\begin{equation}
t_\mathrm{s}=\frac{\rho_\mathrm{d}\,s}{\rho_\mathrm{g}\,C_\mathrm{s}},
\label{EqStopTime}
\end{equation}
where $\rho_\mathrm{g}$ is the gas density.

\citetalias{Stepinski1997} were mostly interested in the radial distribution of
solid particles and their sizes. However, in order to interpret
observations of disks showing evidence of grain settling and growth, as well
as to provide initial conditions for planet formation models, which are
normally axisymmetric situations, one needs to know both the radial and
vertical size and density distributions of grains. The complex interplay
between the drag force, which causes solids to migrate radially and settle
vertically, and the growth process makes a full numerical treatment of this
problem necessary.

\section{Grain growth in a 3D SPH code}
\label{SectSPH}

We have developed a 3D, two-phase (gas+dust) Smoothed Particles Hydrodynamics
(SPH) code to model vertically isothermal, non self-gravitating protoplanetary
disks. The two inter-penetrating phases representing gas and dust interact via 
aerodynamic drag. \citet[hereafter BF05]{BF05}\defcitealias{BF05}{BF05}
describe the code and its limitations, and present the spatial
distribution of dust grains ranging from 1~$\mu$m to 10~m in size resulting
from radial migration and vertical settling. In this work, our aim is to
implement the grain growth algorithm of \citetalias{Stepinski1997} into our
code and see how this simple prescription of grain growth affects the dust
dynamics by comparing with the results of \citetalias{BF05}. While this could
adequately be tested in 2D, our ultimate goal is to study the observational
signatures of grain growth in protoplanetary disks and apply the code to
various non-axisymmetric problems, like disks with embedded planets. Thus we
will be able to extend our previous work on the formation of planetary gaps
in the dust layers of protoplanetary disks \citep{Maddison2007,Fouchet2007}
to include grain growth, and study the stratification of growing dust grains
in disks \citep[see, e.g.,][]{Pinte2007}.

SPH is very well suited to the \citetalias{Stepinski1997} implementation of
grain growth. As we do in our code, \citetalias{Stepinski1997} describe gas
and dust as two fluids, their disk configuration and thermodynamics are very
similar to ours, and our SPH viscosity can be related to a Shakura-Sunyaev
$\alpha$ viscosity \citep{Fouchet2007}. Our test simulations show that the gas
disk is little affected by the evolution of the dust, as assumed by
\citetalias{Stepinski1997}. Therefore, it is straightforward to implement the
\citetalias{Stepinski1997} prescription of grain growth, given by
Eq.~(\ref{EqEvol}), in our code.

Following the work of \citetalias{Stepinski1997}, all the dust particles must
have the same initial size, $s_0$. The size $s$ is then evolved using
Eq.~(\ref{EqEvol}) evaluated at the location of each SPH particle. We assume
that the size $s$ assigned to each SPH particle represents the typical size of
physical dust grains at its location in the disk, at a given time. Again, this
is very similar to the assumption of \citetalias{Stepinski1997} for their local
mean value $s(r,t)$. Contrary to their work, we do not take evaporation into
account: for our disk conditions (see Sect.~\ref{SectSetup}),
\citetalias{Stepinski1997} show that the evaporation radius is located between
1 and 2~AU from the star and that after a million years only a small fraction
of the total solid material is evaporated due to migration. Similar to
\citetalias{Stepinski1997}, our implementation of grain growth does not
include the fragmentation of grains, even though it is likely to play an
important role (see Sect.~\ref{SectConclusion}).

The mass of each SPH particle is kept constant to ensure kinetic energy and
momentum conservation. As $s$ can only increase, this implies that, over time,
SPH dust particles represent fewer but larger physical dust grains. However,
even for km-sized planetesimals, each SPH particle still represents a very
large number of physical particles, maintaining the validity of the numerical
scheme.

\section{Simulations}
\label{SectSimulations}

\subsection{Setup}
\label{SectSetup}

We study grain growth in the typical T~Tauri disk modeled in \citetalias{BF05}
with $M_\mathrm{disk}=0.02\ M_\odot$, composed of 99\% gas and 1\% dust by
mass and orbiting a $1\ M_\odot$ star. The dust grains have an intrinsic
density $\rho_\mathrm{d}=1$~g\,cm$^{-3}$. We choose an initial state for a gas
disk near equilibrium conditions. Following \citet{Hayashi1985}, we take the
parameters of the Minimum Mass Solar Nebula for the radial dependence of the
temperature ($T\propto r^{3/4}$) and initial surface density
($\Sigma\propto r^{-3/2}$). The disk is locally isothermal, i.e. the
temperature follows a radial power law but is vertically constant. The sound
speed then varies as $C_\mathrm{s}\propto r^{-3/8}$ and, given that
$H=C_\mathrm{s}/\Omega_\mathrm{K}$, $H/r$ varies as $r^{1/8}$. The disk is
slightly flared with $H/r=0.05$ at 100~AU.

The smoothing length is computed by $h_i\propto (m_i/\rho_i)^{1/3}$, where
$m_i$ and $\rho_i$ are the mass and density of an SPH particle
\citepalias[see][]{BF05}. The code will not be able to resolve a dust disk for
which the semi-thickness is less than the smoothing length. However,
turbulence in real disks stirs the solid particles and prevents the formation
of such a thin dust layer. Indeed, the dust disk thickness reaches a steady
state when the settling and turbulent mixing are in equilibrium
\citep[see, e.g.,][]{Dubrulle1995,Dullemond2004,Schrapler2004}. Our code does
not yet include turbulent mixing and therefore can not reproduce this steady
state. The SPH artificial viscosity terms are given by
$\alpha_\mathrm{SPH}=0.1$ and $\beta_\mathrm{SPH}=0.0$, which ensures the
corresponding \citet{ShakuraSunyaev1973} viscosity parameter $\alpha\sim0.01$
(as indicated by observations of protoplanetary disks --- see
\citealt{Hartmann1998,King2007}).

We start with 200\,000 gas particles that are distributed radially so as to
retrieve the expected power law for surface density and randomly in the
vertical direction because hydrostatic equilibrium is rapidly reached. The
initial velocity of the gas particles is Keplerian. Starting from this initial
distribution, we allow the gas disk to relax for almost 8\,000 years, which
allows the pressure and artificial viscosity to smooth out the velocity field.
Once the gas disk has relaxed, we then add an equal number of dust particles
on top of the gas particles with the same velocity and allow the system to
evolve.

The disk extends initially from 20 to 300~AU. During the evolution, particles
are removed from the simulation if they migrate inside of 20~AU and are
assumed to be accreted by the star. The outer boundary is free and particles
are only removed if they go beyond 400~AU. The viscous evolution of the disk
is so slow that the gas surface density profile stays almost unchanged
although the disk expands radially up to 400 AU.

We ran a series of simulations with 400,000 SPH particles and initial grain
sizes $s_0$ ranging from 1~$\mu$m to 1~mm, in which the system is evolved for
a total of $10^5$~yr. The results presented in the next sections are time
snapshots of the ongoing disk evolution.

\subsection{Results}
\label{SectResults}

\begin{figure}
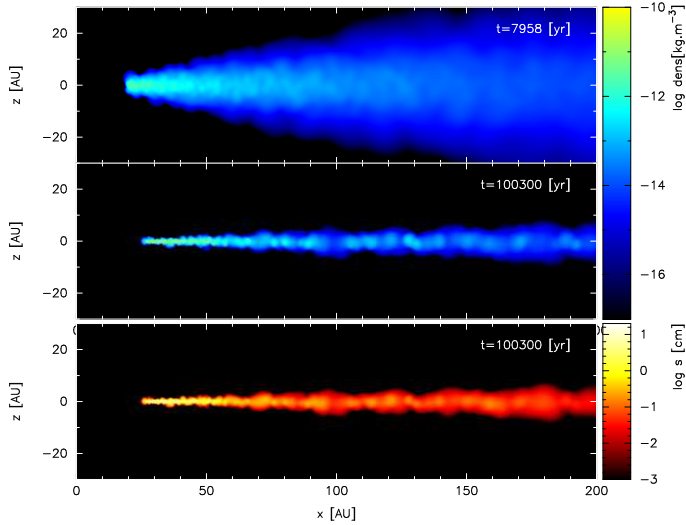

\resizebox{\hsize}{!}{\includegraphics[angle=-90]{rho_begin_end.ps}}
\vspace{-9mm}

\resizebox{\hsize}{!}{\includegraphics[angle=-90]{edge-on2.ps}}
\caption{Initial (top) and final (middle) dust density and final grain size
distribution (bottom) in a meridian plane cut of the disk for $s_0=10\ \mu$m.}
\label{FigXsect}
\end{figure}

\begin{figure}
\resizebox{\hsize}{!}{\includegraphics[angle=-90]{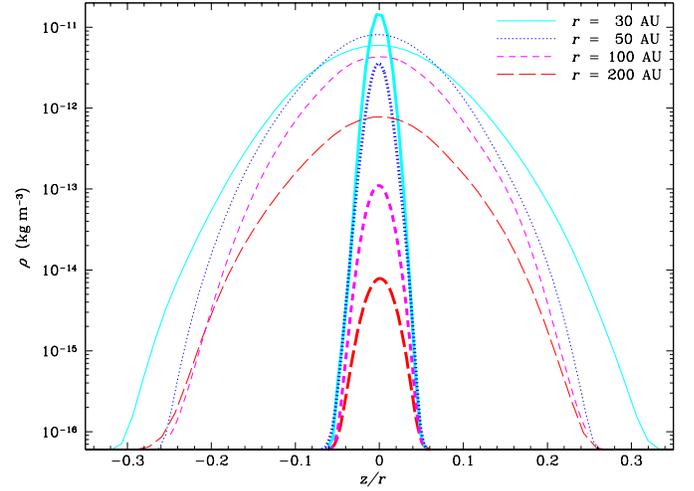}}
\caption{Vertical (azimuthally averaged) profiles of the gas (thin lines) and
dust (thick lines) densities at selected radii in the disk.}
\label{FigDensityVerticalCut}
\end{figure}

Figure~\ref{FigXsect} shows the density of the solid phase at dust injection
(top) and at the end of the simulation (middle), along with the resulting size
distribution (bottom) in a meridian plane cut of the disk for $s_0=10\ \mu$m.
The vertical profiles of the gas and dust densities at the end of the
simulation are shown in more detail in Fig.~\ref{FigDensityVerticalCut}. The
very efficient settling produces a thin dust disk, whereas the gas disk does
not evolve on this timescale.

\begin{figure}
\resizebox{\hsize}{!}{\includegraphics[angle=-90]{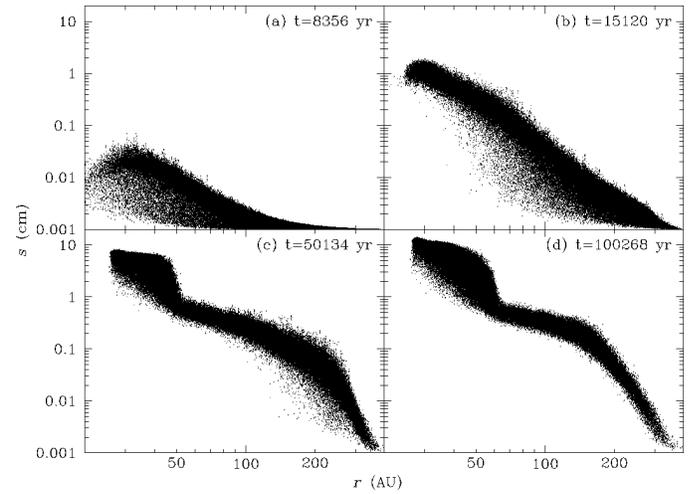}}
\caption{Evolution of the radial grain size distribution for $s_0=10\ \mu$m.}
\label{FigSizeDist}
\end{figure}

Grain growth occurs very quickly: a few hundred years after dust injection,
the innermost grains have almost reached mm sizes (Fig.~\ref{FigSizeDist}a),
and a few thousand years later (Fig.~\ref{FigSizeDist}b), grain growth is
visible over the entire disk, with sizes of a few cm in the inner region.
This fast evolution leads to a radial size distribution showing a regular
increase of grain size with decreasing distance from the star. At later times
(Fig.~\ref{FigSizeDist}c and d), we see a change in the profile shape due to
differences in migration efficiency in different parts of the disk. The
overall distribution then evolves more slowly and shifts to larger sizes while
keeping a roughly constant average slope.

\begin{figure}
\resizebox{\hsize}{!}{\includegraphics[angle=-90]{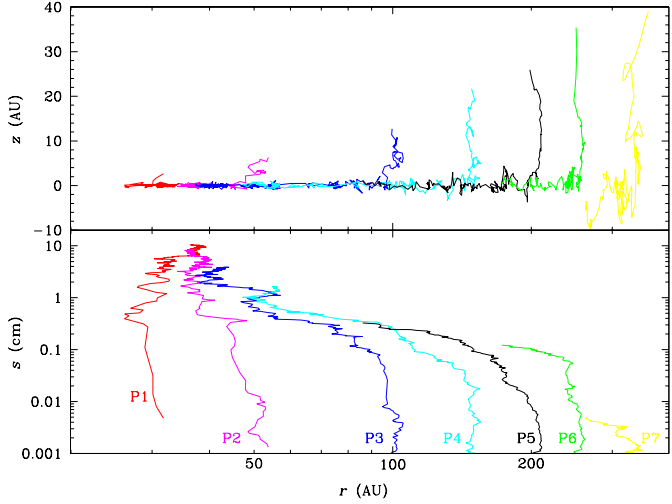}}
\caption{Trajectories of individual particles for $s_0=10\ \mu$m.}
\label{FigTraj}
\end{figure}

In Fig.~\ref{FigTraj}, we show the trajectories of seven individual particles
in the $r-z$ plane for the $s_0=10\ \mu$m case. The top panel shows the
particles settling to the mid-plane followed by their radial migration and the
bottom panel shows the grain growth during this process. We see three stages:
particles grow as they settle to the mid-plane and then start their radial
migration (this behaviour, seen for all particles in Fig.~\ref{FigTraj}, was
also reported by \citet{Haghi2005}), they then rapidly migrate inwards while
growing very little (seen for P2, P3, P4, P5), and finally grow again while
migrating very little (see for P1, P2, P3). The second stage of fast radial
migration corresponds to the low slope seen in the central parts of the disk
in Fig.~\ref{FigSizeDist}c and d: grains spend less time there and accumulate
in the inner disk, where they grow more efficiently due to the higher density.

\begin{figure}
\resizebox{\hsize}{!}{\includegraphics[angle=-90]{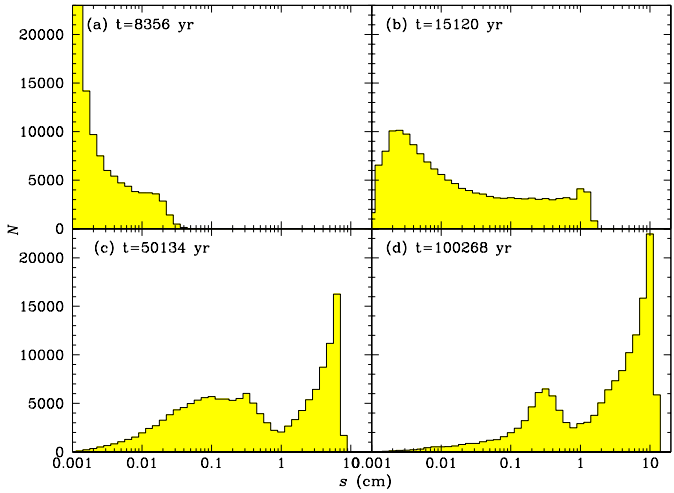}}
\caption{Evolution of the histogram of grain sizes for $s_0=10\ \mu$m.}
\label{FigHisto}
\end{figure}

The final size distribution (Fig.~\ref{FigSizeDist}d) shows a population of
large grains over 10~cm in the inner disk, whereas the grains stay well below
mm sizes in the outer disk. The histogram of final grain sizes for the
$s_0=10\ \mu$m model (Fig.~\ref{FigHisto}d) shows a bimodal distribution. The
first peak corresponds to the end of the first growth stage identified in
Fig.~\ref{FigTraj} where grains reach a size of about 3~mm. The minimum around
8~mm is explained by the rapid migration of the second stage for grains of
that size, which transports them to the very efficient growth region of the
inner disk where they will populate the second peak around 10~cm. As is to be
expected from Eq.~(\ref{EqEvol}), the larger grains are found in the denser
zones (see Fig.~\ref{FigXsect}).

The histograms of grain sizes at different times displayed in
Fig.~\ref{FigHisto} show that the signature of the rapid migration stage as a
minimum around 8~mm appears only a few thousand years after the dust injection
(Fig.~\ref{FigHisto}b), as soon as the larger grains reach that size. It is
visible thoughout the disk evolution, with a remarkably stable position
(Fig.~\ref{FigHisto}c and d).

Changing $s_0$ from 1~$\mu$m to 1~mm only has an effect on the size
distribution in the outer disk, where grains do not grow beyond a few mm. In
the rest of the disk, whatever their initial size, particles quickly reach the
second stage and their subsequent evolution is similar to the $s_0=10\ \mu$m
case, leading to the same grain size distribution.

In this paper, we have restricted our study to disks with a large inner
radius of $r_\mathrm{in}=20$~AU. For lower values, simulations are much slower
due to the smaller drag timestep in the central higher density regions.
Simulations with $r_\mathrm{in}=3$~AU were run for $t=20\,000$~yr and
show the same size distribution outside of 20~AU, with $s$ continuing to
increase as $r$ decreases to 3~AU.

\section{Discussion}
\label{SectDiscussion}

Our results are in good agreement with those of \citetalias{Stepinski1997}.
Indeed, with a disk of comparable mass to ours and an $\alpha$ viscosity
parameter of $10^{-2}$, they found that grain growth is more efficient in the
inner disk, where grains reach sizes of 1~cm at 20~AU after
$3\times10^5$~years. They explain their results by the combined action of both
growth and migration: grains grow, reach sizes where they decouple from the
gas and then migrate, thereby altering the dust surface density and grain
size. Equilibrium occurs when the global flux of migrating particles at a
given radius vanishes. However, they do not resolve the change of slope in the
grain size distribution between 50 and 150~AU. Further, their interpretation
remains qualitative and does not recover the different growth stages we
observe.

In order to interpret our results in a more quantitative way, we compare our
simulations to an analytic expression of the resulting grain sizes derived
under simplifying hypotheses. We first assume that $\bar{v}\ll V_\mathrm{t}$,
i.e.\ that Sc is dominated by the effect of gas-dust coupling, determined by
the value of the non-dimensional stopping time
$T_\mathrm{s}=\Omega_\mathrm{K}\,t_\mathrm{s}$. We introduce the
non-dimensional grain size $S=s/s_\mathrm{opt}$, where
\begin{equation}
s_\mathrm{opt}=\frac{C_\mathrm{s}\,\rho_\mathrm{g}}
{\Omega_\mathrm{K}\,\rho_\mathrm{d}}
\label{EqSopt}
\end{equation}
is the optimal grain size for radial migration
\citep[for details see][]{Fouchet2007}, in order to write
$\mathrm{Sc}\simeq 1+S$. Interestingly, $S=T_\mathrm{s}$ and the three drag
regimes identified by \citetalias{BF05} are defined by $S\ll1$ (dust strongly
coupled to the gas), $S\sim1$ (most efficient dust settling and migration) and
$S\gg1$ (dust decoupled from the gas).

To obtain an analytic solution for Eq.~(\ref{EqEvol}), we then neglect the
temporal variation of all quantities except $s$. Since
\citetalias{Stepinski1997} used constant values for Ro and $\alpha$, the only
remaining variables are $\Omega_\mathrm{K}$, $\rho_\mathrm{g}$,
$\hat{\rho}_\mathrm{d}$ and $C_\mathrm{s}$, which vary both in space and time.
Assuming they are constant when following the evolution of the size of one
particle amounts to assuming that, while it grows, the particle stays in the
same position in the disk (therefore does not settle nor migrate) and that the
disk structure does not evolve with time. While crude, this approximation is
nonetheless useful in understanding the growth process and disentangling its
effects from the complex interaction of those of the drag force. We define the
dimensionless time
\begin{equation}
T=\frac{t}{\tau}+2\sqrt{S_0}\left(1+\frac{S_0}{3}\right),\ \mathrm{with}\ 
\tau=\frac{1}{\sqrt{2^{^3\!/\!_2}\,\mathrm{Ro}\,\alpha}\,\Omega_\mathrm{K}}
\frac{\rho_\mathrm{g}}{\hat{\rho}_\mathrm{d}},
\label{EqTau}
\end{equation}
to rewrite Eq.~(\ref{EqEvol}) as
\begin{equation}
\frac{\mathrm{d}S}{\mathrm{d}T}=\frac{\sqrt{S}}{1+S}.
\label{EqEvolS}
\end{equation}
Its solution is
\begin{equation}
S\!\!=\!\!
\frac{\left(8\!+\!9T^2\!+\!3T\!\sqrt{16\!+\!9T^2}\right)^{^1\!/\!_3}}{2}
\!\!+\!\!
\frac{2}{\left(8\!+\!9T^2\!+\!3T\!\sqrt{16\!+\!9T^2}\right)^{^1\!/\!_3}}\!-\!2.
\label{EqSfinal}
\end{equation}
Three regimes can be isolated:
\begin{equation}
\left|\begin{array}{lclcl}
S\ll1 &\Leftrightarrow&T\ll1&:&S\sim\displaystyle\frac{1}{4}T^2\\[2mm]
S\sim1&\Leftrightarrow&T\sim\displaystyle\frac{8}{3}&:&
S=\displaystyle\frac{T}{2}-\frac{1}{3}\\[2mm]
S\gg1 &\Leftrightarrow&T\gg1&:&S\sim\displaystyle\left(\frac{3}{2}T\right)^{^2\!/\!_3},
\end{array}\right.
\label{EqAsymp}
\end{equation}
showing a fast growth for small sizes and a slower growth for larger sizes,
consistent with what was seen in Fig.~\ref{FigSizeDist}.

Although Eq.~(\ref{EqSfinal}) associates one value of $S$ to a given value of
$T$, the set of normalization quantities ($s_\mathrm{opt},\tau$) defining the
non dimensional $S$ and $T$ is unique for each particle, and is computed with
the disk parameters corresponding to its initial location. The resulting
values of $s$ will therefore be different for all particles.

\begin{figure}
\resizebox{\hsize}{!}{\includegraphics[angle=-90]{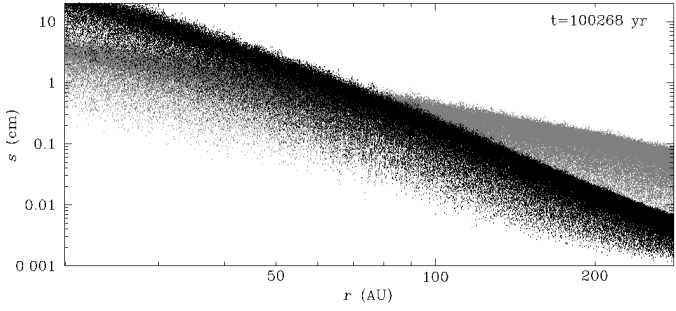}}
\caption{Radial grain size distribution computed from Eq.~(\ref{EqSfinal})
at the end time of our simulations, for $s_0=10\ \mu$m (black) and optimal
size for radial migration $s_\mathrm{opt}$ (grey).}
\label{FigSfinal}
\end{figure}

Starting from the same initial positions of dust particles of size
$s_\mathrm{0}$, this simple model allows us, instead of evolving their size
with the code, to directly compute with Eq.~(\ref{EqSfinal}) the resulting
size (i.e.\ the size computed for the end time of our simulations) of
each particle in the approximation of a fixed position during growth. The
resulting grain size distribution is shown in Fig.~\ref{FigSfinal} for
$s_\mathrm0=10\ \mu$m. As $s$ is proportional to $\rho_\mathrm{g}$ via
$s_\mathrm{opt}$, the observed spread in resulting grain size at a given $r$ is
caused by the spread in gas density for particles at different heights above
the mid-plane. The resulting distribution reproduces the same grain size range
as obtained at the end of the simulation, with decimetric grains in the inner
disk, and has a slope close to that of the distribution in
Fig.~\ref{FigSizeDist}b. Of course, in that approximation,
Eq.~(\ref{EqSfinal}) cannot reproduce the effect of migration and therefore
does not show the changes in slope that it causes, which were seen in
Fig.~\ref{FigSizeDist}c and d.

The understanding of the growth process in the framework of our approximation
which leads to Eq.~(\ref{EqEvolS}) now allows us to reinterpret the results
of grain growth in the presence of vertical settling and radial migration.
For the range of $s_0$ values used in our simulations, $s_0<s_\mathrm{opt}$
over most of the disk (see Fig.~\ref{FigSfinal}), and the grains are initially
in the first regime of fast growth ($S\ll1$) identified in
Sect.~\ref{SectResults}, and found in Eq.~(\ref{EqAsymp}), corresponding to
the strongly coupled drag regime, with very little migration. Outside of
$\sim150$~AU, resulting grain sizes stay below $s_\mathrm{opt}$ in the
simulated timescale and grains do not reach the rapid migration stage. Indeed, 
the growth timescale $\tau\propto\Omega_\mathrm{K}^{-1}\propto r^{^3\!/\!_2}$
is longer in the outer disk. Inside of this radius, sizes do reach
$s_\mathrm{opt}$ (i.e.\ $S\sim1$, corresponding to the intermediate drag
regime) and grains go through a phase of fast inwards migration, explaining
the low slope in the central region of Fig.~\ref{FigSizeDist}d and the stable
location of the minimum in the histograms of grain sizes
(Fig.~\ref{FigHisto}b--d). Indeed, $s_\mathrm{opt}$ depends on variables
describing the gas disk, which does not evolve in the timescale of our
simulations. Once they have grown to sizes greater than $s_\mathrm{opt}$, the
migration efficiency drops as dust starts to decouple from the gas (in the
weak drag regime, defined by $S\gg1$) and grains pile up inside of $\sim50$~AU
and continue to grow there, more slowly.

This result differs from that of \citet{Weidenschilling1977}, who models the
particular case of the Minimum Mass Solar Nebula. With his disk parameters,
he found that the grains with the largest radial migration velocity were
meter-sized and that their consequent survival time in the nebula was much
shorter than the disk lifetime. This led to the persisting
\citep[see, e.g.,][]{Natta2007} idea that growing grains which reached meter
sizes rapidly fall onto the star (or more likely are evaporated in the inner
disk), thus resulting in a potential problem for the planet formation process.
This so-called ``meter-size barrier'' could be overcome if grains could grow
to larger sizes within only a few hundred years, which would be very difficult
according to the current understanding of solid particle aggregation
\citep{Blum2006}. Our simulations show that, in our slightly more massive
nebula representative of many observed T~Tauri disks, growing grains can
survive the fast migration stage (occurring for $\sim8$~mm grains in our disk
conditions) and reach larger sizes with longer survival times, allowing for
the possibility of planet formation inside of $\sim50$~AU.

\section{Conclusion}
\label{SectConclusion}

We have implemented a mechanism able to treat grain growth in protoplanetary
disks via the analytical expression of \citetalias{Stepinski1997} into our
two-phase SPH code. We simulated for the first time the full 3D evolution of
a typical T Tauri disk, following the simultaneous radial migration, vertical
settling, and growth of solid particles. Their interplay is complex: dynamics
affects grain growth by modifying local physical quantities such as density or
relative velocity. Conversely, grain growth also acts on dust dynamics: where
non-growing grains would either stay well mixed with the gas or settle and
migrate according to their sizes \citepalias{BF05}, growing grains will go
through various stages and produce totally different spatial distributions.
They initially grow rapidly as they settle to the mid-plane, then experience
a fast radial migration with little growth through the bulk of the disk, and
finally pile-up in the inner disk where they grow more efficiently. This
results in a bimodal distribution of grain sizes, with the largest grains
found in the denser inner disk, where growth is most efficient. The survival
times of the solid particles are longer than previously found, which has an
implication on planet formation.

We find that grains grow very quickly: they reach decimetric size in $10^5$~yr.
This is in general agreement with the results of \citet{DullemondDominik2005}
where shattering is neglected. They used the different approach of solving the
Smoluchowski equation to study the growth of settling, but non-migrating, dust
and in $10^5$~yr formed grains of maximum sizes ranging from  1~cm to over 1~m
depending on their model parameters. Similar to them, we also find that the
small grains only survive in the very outer disk and are depleted too rapidly
elsewhere to be consistent with infrared observations of disks, highlighting
the importance of shattering.

In order to compute synthetic images from our simulations and compare them to
the observations, one would have to assume that the collisional cascade
resulting from the inclusion of shattering would produce a whole particle
distribution from the maximum size at a given radius shown in
Fig.~\ref{FigSizeDist}d down to sub-$\mu$m size, described by a quasi-steady
power law as argued by \citet{Garaud2007}.

The method we used to treat grain growth can easily be applied to other
analytical prescriptions.  The development of a more detailed model is
necessary for a realistic description of grain growth in protoplanetary disks.
In addition to shattering, one needs to take into account other processes such
as microscopic interactions between the grains, kinetic energy dissipation and
grain porosity. This is the subject of a forthcoming paper.

\begin{acknowledgements}
We thank Yann Alibert for suggesting the use of the \cite{Stepinski1997}
approach. This research was partially supported by the Programme National de
Physique Stellaire of CNRS/INSU, France, the Programme International de
Coop\'eration Scientifique (PICS) France-Australia in Astrophysics (Formation
and Evolution of Structures), and the Swinburne University Research Development
Grant Scheme. Simulations presented in this work were run on the Swinburne
Supercomputer\footnote{\tt http://astronomy.swin.edu.au/supercomputing/}.
Images in Fig.~\ref{FigXsect} were made with SPLASH \citep{Price2007}.
\end{acknowledgements}

\bibliographystyle{aa}
\bibliography{bibliodisks}

\end{document}